# Emergence, self-organization and network efficiency in gigantic termite-nest-networks build using simple rules

Diego Griffon[1], Carmen Andara[2], Klaus Jaffe[3]
[1] PFG en Agroecología, Universidad Bolivariana de Venezuela. Caracas, Venezuela.
[2] Departamento de Biología. Facultad de Ciencias y Tecnología. Universidad de Carabobo. Valencia, Venezuela.
[3] Departamento de Biología de Organismos. Universidad Simón Bolívar. Caracas, Venezuela

**Termites, like many social insects, build nests of complex architecture. These constructions have been proposed to optimize different structural features. Here we describe the nest network of the termite *Nasutitermes ephratae*, which is among the largest nest-network reported for termites and show that it optimizes diverse parameters defining the network architecture. The network structure avoids multiple crossing of galleries and minimizes the overlap of foraging territories. Thus, these termites are able to minimize the number of galleries they build, while maximizing the foraging area available at the nest mounds. We present a simple computer algorithm that reproduces the basics characteristics of this termite nest network, showing that simple rules can produce complex architectural designs efficiently.**

Termites often built nest-networks where nests are connected to each other by covered galleries (Holt & Easey 1985; Vasconcellos & Gomes 2006; Husseneder et al. 1998). These networks can achieve a complex architecture and have been proposed to optimize different structural features (Korb & Linsenmair 1998; Cole 1994; Farji-Brener 2000). The large-scale structure of these nest-networks is an emergent propriety resulting from a set of parallel distributed decisions made by individual insects (Sole & Goodwin 2000; Buhl at al.2004). Its final topology is generated by the extension, branching, and intersection of the growing parts of the network (Buhl at al.2004). Here we study how efficient these structures are, taking into account that no blueprint of these nest-networks can possibly exist, as their construction are beautiful examples of self-organization (Sole & Goodwin 2000). In addition to the efficiency concept we use a biologically more meaningful concept such as *satisficing* strategies (Simon 1955) in order to describe complex nest building rules among termites.

The termites studied are located in Canaima National Park, Venezuela. They build nest-networks made of superficial nests inter-connected by covered galleries (Figure 1). The largest network observed had 645 nests covering a surface of 2100 $m^2$. Similarly large networks had been reported before for *Reticulitermes flavipes* (2365 $m^2$: Su at al.1993) and *Coptotermes formosanus* (3571 $m^2$: Su & Scheffran 1988), although using different techniques (Hu, Zhong & Guo 2006) to the one we used. In the case of *N. ephratae,* there are two types of nests in the termite network: big ones (approximately 60 cm of diameter) called "mains" and small ones (approximately of 15 cm of diameter) called "satellites".

For the compilation of the network information, each individual nest was numbered in field and its galleries were registered on an adjacency matrix (Albert & Barabasi 2002; Dorogovtsev & Mendes 2002; Newman 2003). In order to calculate descriptive features of complex systems we computed its graph fractal dimension or GFD which is a measure of the dimensionality of the spatial structure, the minimum, spanning tree or MST which gives an estimate of the efficiency in the use of connections and the foraging area. For the calculation of the GFD we constructed a geodesic matrix using the Pajek software (Batagelj & Mrvar2003). The GFD was calculated on the geodesic matrix according to refs

10. For the calculation of the MST of each network the Grin software (Pechenkin 2007) was used. For the calculation of the foraging area the ImageJ software (Abramoff, Magelhaes & Ram 2004) was used.

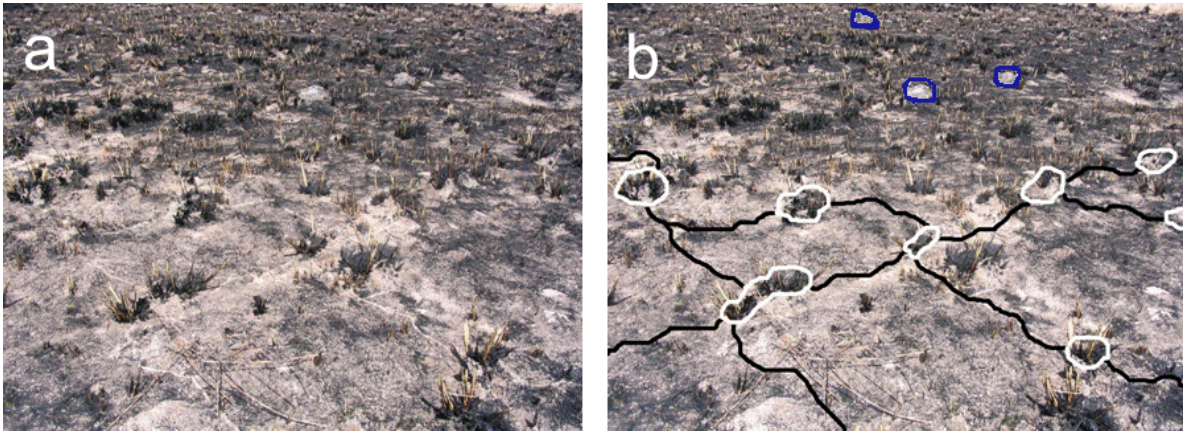

**Figure 1**. Network of nests and galleries of *Nasutitermes ephratae* after a natural fire made its structure visible. 1a, Fragment of one network. 1b, A detail of the same fragment but with elements of the network made more visible; blue line: main nests; white line: satellite nests; black line: galleries.

We studied 3 networks which we called very big, containing 16 main nests, 629 satellite nests and 707 galleries; big, with 3 main nests, 75 satellite nests and 88 galleries; and small, with 3 main nests, 67 satellite nests and 83 galleries. The networks were almost completely covered by dense herbaceous vegetation. The data was collected after a natural fire made the networks visible. The networks are located on herbaceous savannas next to the Parupa Scientific Station (66º 36' 55" E   6 º 34' 30" N).

These termite nest networks showed a noticeable bias towards nests connected by two galleries, forming chains that appeared like pearl collars (Figure 2a). The connectivity distribution (*i.e.*, histogram of the number of galleries ($k$) per nest) of the three networks showed a bias towards connectivity = 2 (Figure 2b). In all cases, the maximum number of galleries per nest was 8 (*Kmax*). Not all galleries might be in use as some might have been abandoned because they were useless for the colony or inefficient.

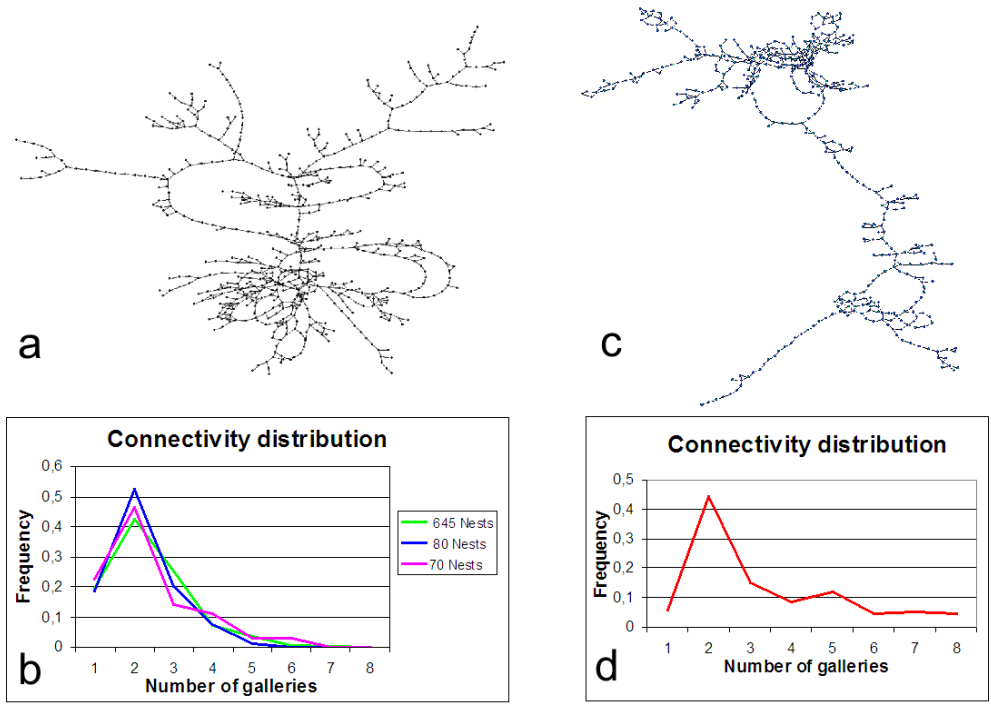

**Figure 2**. Nest networks and connectivity distribution. 2**a**, Representation of network with 645 nests. 2**b**, connectivity distribution of the three studied networks (blue line: 645 nests, purple line: 80 nests, red line: 70 nests). 2**c**, A simulated network. 2**d**, Connectivity distribution of the simulated network.

A mayor restriction of these networks is that, since they are built on the surface of the ground, they have to be 2D networks. These networks can easily develop inefficient structures, like multiple crossing of galleries in a specific area (Gorman & Kulkarni, R. 2004). Networks that avoid these inefficient structures are called Planars (*i.e.*, a network that can be draw in 2D without crossings its links, see Gorman & Kulkarni 2004; Gastner & Newman 2006; Gastner & Newman 2006). A network is considered to be planar if it has a graph fractal dimension (GFD) smaller than two (Gastner & Newman 2006a). The GFD of the termite nest-networks on average were of $1.6 \pm 0.03$ (calculated according to the procedure propose by Gastner & Newman 2006a), showing that they were build avoiding multiple crossing of galleries. A very similar number was obtained for *C. formosanus* and *R. flavipes* when studying subterranean galleries (Puche and Su), suggesting that this might be a general feature of termite gallery networks.

In these networks, the angles between consecutive galleries were not random. Sixty percent of the angles between pairs of galleries were in the range of 60º to 120º, the second most common where within the range of 160º to 180º (25% of the angles). It is interesting to notice, that if we assume foraging territories (i.e. uncovered foraging areas at the exit of a trial) of constant size, those angles avoid the overlap of foraging territories (Figure 3). Thus, the angles between galleries optimize foraging activity.

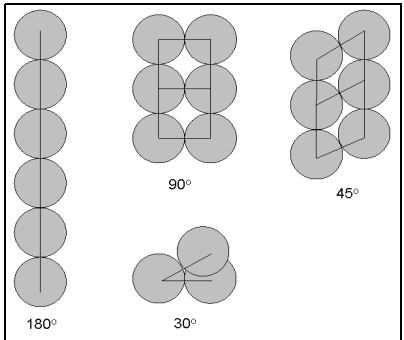

**Figure 3**. Examples of angles between galleries (lines). Foraging territories areas of constant size are represented in circles. Angles of 180º or 90º (±45) avoid overlapping of foraging territories. Others angles produce overlap as in territories branching at 30º.

The construction of the network involves an energy cost. A way to optimize energy usage is to build the network using the minimum number of galleries possible. A graph that connects all its nodes using the minimum number of links is known as a minimum spanning tree or MST (Buhl et al. 2004; Gastner & Newman 2006b; Schweitzer et al. 1998. Any network can be transform into a MST (Buhl at al.2004). The MST obtained from this transformation represented the optimum way to connect the nodes of the original network (using the minimum number links as in Buhl at al.2004). We compared the number of galleries in the termite nest-networks with the number of links in a MST obtained from each network (see Sole & Godwin 2000), and found that the relation: termite-network gallery number / MST links number, is on average $0.85 \pm 0.12$. This indicates, in terms of gallery usage, that these networks are close to the optimum (=1).

We also studied the efficiency of the network in terms of the foraging area available to the termites, that is, the area on which the termites can potentially collect grass. These termites forage on open land every night to a maximum distance of 0.5 m from the nearest nest structure, although covered galleries show termite activity also during the day when opened forcefully. Based on this information, we delineate on a map drawn to scale, a perimeter that represent the maximum distance that termites can forage. Then we calculated the area generated between the network structure and the foraging perimeter (green area in Figure 6a). We found that the network architecture assured an almost optimal use of the area around the main nests. If we draw a circle with a diameter of 5,5 m around those nests, we observe that on average termites can access 77,8% ($\pm$ 5) of the available foraging area (Figure 6b).

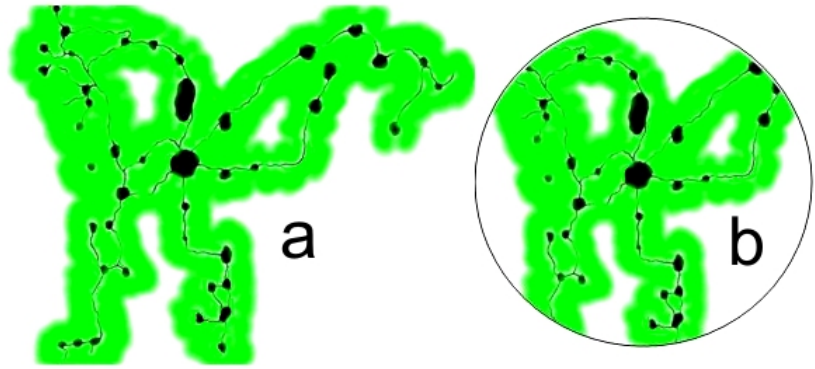

**Figure 4**. **a**, Foraging area accessible from the termite nest-network (black: termite nest-network; green: potential foraging area). **b**, Area around a main nest (the circle encloses an area of 24 m$^2$).

Finally, we searched for simple computational algorithms that could generate networks of similar characteristics to those build by the termites. We found that similar networks can be obtained by executing the steps of a model that assumes a constant gallery length and a homogenous environment (these two restrictions are compatible with the empirical data).

*(1)* The model starts with one initial nest built on the center of a grid.

*(2)* In the neighborhood of the previous nest (a Moore type neighborhood of 8 cells) a new nest is formed with a homogenous probability for all cells. The new nest is linked to the old one by a gallery.

*(3)* Then, in the neighborhood of the new nest, another nest is founded in the same way described in *(2)*.

*(4)* Steps *(2)* and *(3)* are repeated *l* times. In this way, a series of new galleries and nests are constructed by means of simulated random walks of length *l*, were *l* is the number of times a new nest is build on the simulated random walk.

*(5)* After finished a random walk, according to a probability *p,* the initials or finals nests of the walks already made can give origin to another walk (with the restriction that the number of times that a walk begins on a nest, must be smaller or equal to *Kmax*).

*(6)* Steps *(2)* to *(5)* are repeated for *i* iterations.

This simple model, under an ample range of values of parameters *l*, *p* and *I*, reproduces the basic characteristics of the termite network (Figure 2d) and its connectivity distribution (Figure 2c).

Herbert Simon (1955) defined a *satisficing* strategy as one which attempts to meet criteria for adequacy, rather than to identify an optimal solution. This seems to be the case of the termite nest-networks architectures, which are satisficing in terms of the compromise between network construction energy costs and potential foraging area generate, all this achieved with an architecture that can be build with simple behavioral rules. Other simulations (Lee Baedunias & Su 2008) based on simple rules were also able to reproduce termite foraging behavior of a different kind, suggesting that termite behavior might be based on a limited number of simple rules. Knowing these rules allows now for experimental laboratory studies in order to falsify or expand our hypothesis.

**Acknowledgements:** We thank Dayaleth Alfonzo, Solange Issa, Augusto Andara, Roberto Cipriani, Glenda Briceño, Javier Rodriguez, Rico Rodriguez and Parupa Scientific Station personal for their help in the realization of this research.